\documentclass[%
 reprint,
 amsmath,amssymb,
 aps,
]{revtex4-1}

\usepackage{graphicx}
\usepackage{dcolumn}
\usepackage{bm}
\usepackage{hyperref}

\usepackage{color}

\begin{document}

\preprint{APS/123-QED}

\title{Efficient Exploration of Multi-Modal Posterior Distributions}

\author{Yi-Ming Hu}
 \email{y.hu.1@research.gla.ac.uk}
\author{Martin Hendry}%
\author{Ik Siong Heng}
\affiliation{%
 School of Physics and Astronomy, University of Glasgow, Glasgow, G12 8QQ, UK\\
}%

\date{\today}

\begin{abstract}
The Markov Chain Monte Carlo (MCMC) algorithm is a widely recognised as an efficient method for sampling a specified posterior distribution.
However, when the posterior is multi-modal, conventional MCMC algorithms either tend to become stuck in one local mode, become non-Markovian or require an excessively long time to explore the global properties of the distribution.
We propose a novel variant of MCMC, \emph{mixed MCMC}, which exploits a specially designed proposal density to allow the generation candidate points from any of a number of different modes.
This new method is efficient by design, and is strictly Markovian.
We present our method and apply it to a toy model inference problem to demonstrate its validity.
\end{abstract}

\maketitle


\section{\label{sec:Intr}Introduction}

Bayesian inference methods have been applied to an increasingly wide range of data analysis problems in the physical sciences -- particularly problems requiring \emph{parameter estimation} and \emph{model selection} (see, for example, \cite{gregory05}\cite{sivia06}\cite{rover07}\cite{loredo13}). This growing popularity of the Bayesian approach has been driven by the ready availability of increasingly powerful computational facilities, allied with significant advances in the data analysis methodology and algorithms that have been developed.  These twin drivers have permitted the application of Bayesian methods to data sets of a size, complexity and dimensionality that until recently would have rendered them intractable.  

One of the main targets of Bayesian inference is to estimate the posterior distribution of desired parameters. To estimate posterior distributions, one naive solution is to use an exhaustive algorithm to calculate the posterior over a dense grid of points in the parameter space. Such brute-force methods will have little or no practical value when dealing with medium-to-high dimensional problems since the computational burden will be prohibitively high. For such problems the ability to concentrate sampling in regions where the posterior probability is high is very important if we are to implement Bayesian inference methods efficiently.

Methods such as \emph{Markov Chain Monte Carlo (MCMC)} and \emph{Nested Sampling} are well tailored to explore the posterior distribution over high dimensional parameter spaces.  While the computational cost of brute-force methods increases exponentially with the dimension, MCMC usually only grows linearly with dimension \cite{allison13}\cite{veitch10}.

Generally, the method of MCMC works well so long as the posterior surface is sufficiently smooth. However, when the posterior distribution has a complicated structure, MCMC will become inefficient.  For example, MCMC samplers are known to get ``caught" in a local mode of the posterior, and unable to jump out and explore any other isolated modes in the parameter space \cite{kirkpatrick83}\cite{neal96}.
So a lot of methods have been proposed to make the MCMC sampling more efficient ({\em e.g.\/} \cite{tierney99}\cite{trias09}\cite{mira01})

In this work, we propose a novel method called \emph{mixed MCMC\/} to deal with such issues. The conventional MCMC algorithm is robust for exploring the detailed structure of the posterior surface, and we want to retain that property while enabling some global ``communication" between different regions of the parameter space so that the sampler can make jumps between those regions without requiring a very long exploration time. 

This paper is structured as follows.
In section \ref{sec:Met}, we introduced the general realisation of MCMC and discussed how the MCMC can be used as a Bayesian tool.
In section \ref{sec:mix}, we discuss the main difference of mixed MCMC and conventional MCMC and give the pseudo code for its realisation.
We apply the method to a toy model as illustrated in section \ref{sec:toy}.
Finally in section \ref{sec:dis}, we summarise the motivation and properties of mixed MCMC, and discussed possible extension to it.

\section{Method}\label{sec:Met}

The basic principles of an MCMC algorithm are simply stated. The algorithm sets out to sample a chain of points in the parameter space and at the $i^{th}$ iteration (i.e. after $i-1$ points have already been sampled) a \emph{candidate\/} point $\boldsymbol{\theta^*}$ is randomly sampled from some specified proposal distribution, based solely on the position of the \emph{previous point\/} in the chain $\boldsymbol{\theta^{(i-1)}}$. The corresponding posterior for this candidate point is calculated, and compared with the posterior at $\boldsymbol{\theta^{(i-1)}}$.  If the value of the posterior at the candidate point is larger than that of the previous point, the candidate point is accepted as the next point in the chain.  Otherwise, the candidate is accepted only with a certain \emph{acceptance probability} (see next section).   One finds therefore that the sampling will generally proceed ``uphill" -- i.e. to regions of the parameter space where the value of the posterior is larger -- while sometimes it can also go ``downhill" to regions where the posterior takes on lower values.  The precise form of the acceptance probability achieves what is termed \emph{detailed balance\/}, which ensures that the chain of sampled points is indeed a random sample drawn from the desired posterior distribution.  This method can thus be used to efficiently explore the posterior distribution, avoiding the need for a global optimization via an exhaustive grid search.\cite{gregory05}\cite{sivia06}\cite{rover07}\cite{loredo13}

\subsection{Markov Chain Monte Carlo}

Interested reader is refered to \cite{gregory05} and \cite{sivia06} for detailed discussion of Bayesian Inference.
Hereafter, we define the posterior $f(\boldsymbol{\theta})=p(\boldsymbol{\theta}|D,I)$, the prior $\pi(\boldsymbol{\theta})=p(\boldsymbol{\theta},I)$ and likelihood $\ell(\boldsymbol{\theta})=p(D|\boldsymbol{\theta},I)$, where $\boldsymbol{\theta}$ is the parameter set, $D$ is the data and $I$ is the information.

The simplest form of Markov Chain Monte Carlo (MCMC) is known as the Metropolis algorithm, which can be achieved by the following steps \cite{gregory05}\cite{sivia06}\cite{rover07}\cite{press07}.

\begin{enumerate}
\item Arbitrarily choose a starting point $\boldsymbol{\theta}^{(0)}$ that satisfies $f(\boldsymbol{\theta}^{(0)})>0$, and a symmetric proposal distribution $J(\boldsymbol{\theta}_a|\boldsymbol{\theta}_b)$. Set step index \emph{i}=0.

\item Increment \emph{i} by 1.
\item Randomly propose a new parameter set $\boldsymbol{\theta}^*$ by sampling from $J(\cdot|\boldsymbol{\theta}^{(i-1)})$.

\item Calculate the \emph{Metropolis ratio\/} given by
	\begin{equation}\label{eq:accp_prob}
		 r=\frac{f(\boldsymbol{\theta}^*)}{f(\boldsymbol{\theta}^{(i-1)})}
	\end{equation}

\item Accept the proposed parameter set $\boldsymbol{\theta}^*$ with \emph{acceptance probability} 
\begin{equation}\label{eq:accpt}
\alpha(\boldsymbol{\theta}^{(i-1)},\boldsymbol{\theta}^*)\triangleq\min(1,r)
\end{equation}

	If $r\geq1$, then the candidate is accepted, so the new point is $\boldsymbol{\theta}^{(i)}=\boldsymbol{\theta}^*$.

	If $r<1$, draw a random number $rand$ from a uniform distribution $U[0,1]$, and if $rand<r$, then set 
	$\boldsymbol{\theta}^{(i)}=\boldsymbol{\theta}^*$; otherwise set
	 $\boldsymbol{\theta}^{(i)}=\boldsymbol{\theta}^{(i-1)}$.

\end{enumerate}
Step 2-5 are repeated until a large enough number of points have been sampled. This termination could be controlled by a preset number, or by monitoring the samples' distribution and check if it's sufficiently stable.\cite{gelman92} The beginning peorid, which is generaly called as ``burn-in" stage, is discarded to prevent the influence of the arbitrary choice of starting point $\boldsymbol{\theta}^{(0)}$.

The Metropolis-Hastings (M-H) algorithm is a more general form of the Metropolis algorithm. 
In the Metropolis algorithm, the proposal distribution is symmetric, that is $J(\boldsymbol{\theta}_a|\boldsymbol{\theta}_b)=J(\boldsymbol{\theta}_b|\boldsymbol{\theta}_a)$, but this condition is not necessary. 
In the M-H algorithm we relax this symmetric condition, so that equation (\ref{eq:accp_prob}) should be modified as follows
\begin{equation}\label{M-H}
	 r=\frac{f(\boldsymbol{\theta}^*)J(\boldsymbol{\theta}^{(i-1)}|\boldsymbol{\theta}^*)}{f(\boldsymbol{\theta}^{(i-1)})
	 J(\boldsymbol{\theta}^*|\boldsymbol{\theta}^{(i-1)})}.
 \end{equation}

It is clear that when the proposal distribution is symmetric, equation (\ref{M-H}) is identical to equation (\ref{eq:accp_prob}).

It can be shown that the number density of the sampled points will represent a sample from the posterior distribution~\cite{gregory05}. Thus estimation of the parameter(s) that characterise the posterior distribution becomes possible with a sufficiently large number of sampling points.
 
\subsection{Convergence of MCMC}

We can estimate the posterior distribution from a histogram of the values sampled by our MCMC chain, and the mean of the parameters can be estimated trivially as 
$$ \frac{1}{N}\sum_{i=1}^N \boldsymbol{\theta}^{(i)}$$
Any realization of MCMC is guaranteed to be converged if it satisfies the requirement of \emph{detailed balance}.\cite{gregory05}
  \begin{equation}\label{detail_balance}
	   f(\boldsymbol{\theta}^{(i)})J(\boldsymbol{\theta}^{(i-1)}|\boldsymbol{\theta}_t)=f(\boldsymbol{\theta}^{(i-1)})
	   J(\boldsymbol{\theta}^{(i)}|\boldsymbol{\theta}^{(i-1)})
   \end{equation}

The concept of detailed balance in thermodynamics can help us to understand this requirement for the convergence of the MCMC chain. In thermodynamics, we can define the probability of a particle to be in state $\boldsymbol{\theta}^{(i)}$ as $f(\boldsymbol{\theta}^{(i)})$, and the probability of the particle to jump to state $\boldsymbol{\theta}^{(i-1)}$ as $J(\boldsymbol{\theta}^{(i-1)}|\boldsymbol{\theta}^{(i)})$.  Detailed balance requires that, after a sufficiently long timescale, the probability for a particle to jump from state $\boldsymbol{\theta}^{(i)}$ to state $\boldsymbol{\theta}^{(i-1)}$ should be exactly the same as the probability to jump from state $\boldsymbol{\theta}^{(i-1)}$ to state $\boldsymbol{\theta}^{(i)}$. 

We need to notice that detailed balance is a stronger requirement than convergence, in the sense that a Markovian Chain that is not in detailed balanced may still converge to the target distribution.\cite{foreman13}

\section{mixed MCMC}\label{sec:mix}

If the starting point and/or the proposal density is not properly chosen, the MCMC sampler might become stuck in a local mode, and will not be able to appropriately explore the whole parameter space.  This might introduce a statistical bias in the parameter estimation carried out by MCMC, particularly when the target distribution is multi-modal. Thus motivates the realisation of mixed MCMC as a really Markovain realisation of MCMC that can sample posterior efficiently.\cite{neal96}\cite{gelman92}\cite{foreman13}

Here we propose a novel method which we term \emph{mixed MCMC\/} to perform Bayesian inference on multi-modal posterior distributions.  This method can allow the sampler to communicate between different local maxima, so that the sampler will be able to represent local peaks, as well as to explore the global structure.  As noted previously, our method requires some limited information about the location of the multiple modes before sampling.  In many cases, however, we will have at least some rough prior knowledge about the posterior, and we can use this information to guide the sampler. Even in the absence of such prior knowledge, other existing global sampling methods can be tailored for this purpose speed up this process.\cite{neal96}\cite{feroz13}\cite{crowder06}

\subsection{Algorithm}

The main difference between the algorithm for mixed MCMC and the conventional MCMC algorithms simply roots in the use of a novel form of proposal density. The sampler should be able to generate candidates from different sub-regions, while proper choice of Metropolis ratio will ensure that the sampling between those different sub-regions satisfies detailed balance.

Suppose, as a result of existing prior knowledge, or with the help of some other global sampling method, we have some information about the posterior distribution that is sufficient to identify the existence and the rough location of the several modes in posterior distribution, where the location of the $t^{th}$ mode is labeled as $\boldsymbol{\theta^0_t}$.  We can then divide the parameter space into several distinct \emph{sub-regions} each of which we assume contains a single mode of the posterior \cite{gregory05}\cite{gregory07}.

We should bare in mind that this method is designed for multi-modal posterior, thus the proposal density should be designed in a way that it can propose new candidates in all posterior modes. Thus we assign to the $t^{th}$ sub-region what we term a  \emph{picking up probability\/}, $p_{t}$, which determines the probability to get a new candidate in the $t^{th}$ sub-region.  Ideally, this probability should be the same as the marginal likelihood (also known as the evidence) within the sub-region -- i.e. the probability that the candidate point lies within that sub-region.  Note also that the picking up probability should satisfy the normalisation requirement
\emph{ $\sum\limits_t p_{t}=1$}.   At the same time it will maximise the efficiency of our approach if $p_{t}\propto\int_{V_t}f(\boldsymbol{\theta})d\boldsymbol{\theta}$, where $V_t$ is the volume of the $t^{th}$ sub-region of the parameter space. 

Suppose we decide to generate a candidate point in the $t^{th}$ sub-region, while the current (i.e. most recently updated) point $\boldsymbol{\theta^{(i-1)}}$ is located in the $s^{th}$ sub-region.  Then a normalised multivariate distribution (most conveniently taken to be a Gaussian) centering around the point $\boldsymbol{\theta^0_{t}} - \boldsymbol{\theta^0_{s}} +\boldsymbol{\theta^{(i-1)}}$ is used as proposal density, and a candidate is drawn from this distribution.  After calculating the value of the posterior at this candidate point, and then computing the Metropolis ratio, $r$, in the usual way, we can decide to accept the candidate point with the acceptance probability $\alpha$ as before. 

In more detail our mixed MCMC algorithm can be illustrated with the following pseudo-code.
\begin{enumerate}
\item Obtain some rough approximation to the posterior distribution using other methods. 
Identify \emph{m} modes in the parameter space, and estimate their central locations given by $\boldsymbol{\theta^0_{t}}$. 
\item Set $p_{t}$ to be the picking up probability, defined as proportional to the volume of the $t^{th}$ sub-region, with $\sum_{t} p_t=1$.  Set step label $i=0$
\item Randomly pick a starting point, $\boldsymbol{\theta^{(i)}}$.
\item \emph{while}(not converged)
		\begin{enumerate}
			\item Set i=i+1
			\item Randomly pick a sub-region number $t$ with probability $p_{t}$ and assign $s$ to be the current sub-region index. $s,~t\in\{1,...,m\}$ where m is the number of all sub-regions.
			\item Generate the candidate point $\boldsymbol{\theta^*}=\boldsymbol{\theta^{(i-1)}}+\boldsymbol{\theta^0_{t}}-\boldsymbol{\theta^0_{s}}+\boldsymbol{\delta \theta}$ drawn from the proposal density $\sim 
			J(\cdot|\boldsymbol{\theta^{(i)}})$
			\item Calculate the Metropolis ratio $r$ based on the candidate and the previous point,  $r=\rm \frac{f(\boldsymbol{\theta^{*}})p_t}{f(\boldsymbol{\theta^{(i-1)}})p_s}$. 
			\item Generate a random number ${\rm rand}\sim U[0,1]$.
			\item Accept the proposed parameter set $\boldsymbol{\theta}^*$ with \emph{acceptance probability} $\alpha=\min(1,r)$ as follows:

			\emph{if}($r>{\rm rand}$),~~~~
			update, $\boldsymbol{\theta^{(i)}}=\boldsymbol{\theta^{*}}$
			\\ \emph{else}~~~~~~~~~~~~~~~~~~$\boldsymbol{\theta^{(i)}}=\boldsymbol{\theta^{(i-1)}}$
		\end{enumerate}

\end{enumerate}

The mixed MCMC algorithm set out above is strictly Markovian, and detailed balance is achieved by construction. Thus the number of points sampled in given sub-region should provide an estimate of the local evidence.  As noted above, in order to maximise the efficiency of the algorithm the picking up probability $p_t$ should better be proportional to the local evidence.

Also, we can notice that when the proposed point and the previous point are located in the same sub-region, then the algorithm reduces to the conventional M-H algorithm, which further verifies its validity.

\section{Toy Model}\label{sec:toy}
We demonstrate our mixed MCMC algorithm using a simple toy model. On a two dimensional $x-y$ parameter space, we considered a posterior distribution is the sum of a pair of well-separated bivariate normal distributions.
Where the parameters are $x$ and $y$, and the two artificial posterior modes locate in $(\mu^x_1,\mu^y_1)$ and $(\mu^x_2,\mu^y_2)$, each mode can be described by a bivariate normal distribution with a diagonal covariance matrix, where standard deviation in each direction is $\sigma^x_1,\sigma^y_1$ for the first mode and $\sigma^x_2, \sigma^y_2$ for the second mode.
The form of this posterior is, therefore:
\begin{equation}\label{posterior}
\begin{split}
f(x,y) = C_1\exp[-\frac{(x-\mu^x_1)^2}{2(\sigma^x_1)^2}-\frac{(y-\mu^y_1)^2}{2(\sigma^y_1)^2}]
\\+C_2\exp[-\frac{(x-\mu^x_2)^2}{2(\sigma^x_2)^2}-\frac{(y-\mu^y_2)^2}{2(\sigma^y_2)^2}].
\end{split}
\end{equation}
The coefficients $C_1$ and $C_2$ allow the two modes to differ in height, and when integrated over the entire parameter space the normalization condition implies that 
\begin{equation}\label{post_norm}
\int_{x,y}\mathrm{d}x~\mathrm{d}y~ f(x,y) = 1.
\end{equation}
For simplicity, we chose $\mu_1^x=-\mu_2^x=-3$, $\sigma_1^x=\sigma_2^x=0.1$, $\mu_y^1=\mu_y^2=0$, $\sigma_1^y=\sigma_2^y=0.1$ and the ratio of two coefficients $C_1:C_2$ is kept as $1:3$.

In this toy model test, we only concentrate on the validity of the mixed MCMC method, and do not consider in detail other factors such as its efficiency or generality.  Thus, we assume prior knowledge of the separated structure of the posterior distribution. Given this assumption, it is possible to analytically calculate the $\Delta\chi^2$ value that corresponds to the contour within which a certain fraction of the entire volume of the posterior is located, thus providing us with an exact theoretical reference result with which to compare.
For a one-dimensional Gaussian distribution, the $1\sigma$, $2\sigma$ and $3\sigma$ credible regions correspond to $68.27\%,~    95.45\%$ and $    99.73\%$ of the cumulative probability function (CDF) respectively. Thus, it is convenient to consider for our toy model posterior the $\Delta \chi^2$ values that correspond to the $68.27\%,~    95.45\%$ and $   99.73\%$ of the CDF, and compare it with the sample estimates obtained from application of our mixed MCMC algorithm.

Under the Gaussian assumption, $\Delta \chi^2 = -\Delta 2\log(\cal L)$, so in the general case we will compare the value of $2\Delta\log(\cal L)$ with its theoretical evaluation. Details of this theoretical calculation can be found in the Appendix.  The posterior of the toy model can be taken as two independent bivariant Gaussians, each with a diagonal covariance matrix.

We generated a chain with $10^5$ points. The proposal density was set to be a bivariate Gaussian 
distribution in addition to the probabilistic shift between sub-regions, with covariance matrix equal to the  identity matrix multiplied by $\sigma=0.1$.

For this particular toy model, theoretically the corresponding $68.27\%,~    95.45\%$ and $   99.73\%$ credible regions should have values of $2\Delta\log(\cal L)$ equal to $3.11,~6.99$ and $12.64$. 
A typical realisation gives result as $3.14,~7.04$ and $12.24$, and the numbers of points sampled in the two sub-regions are 2559 and 7441, which is consistent with the $1:3$ ratio assumed for the coefficients $C_1$ and $C_2$.

\begin{figure}[htbp]
\centering
\includegraphics[width=0.5\textwidth]{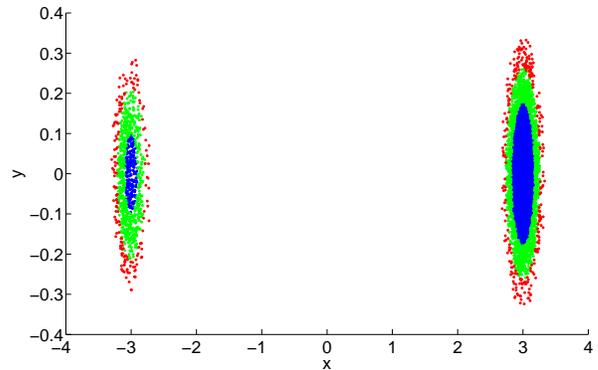}
\caption{Example realisation of mixed MCMC applied to the toy model posterior distribution described in the text.  Here blue, green and red colour points represent subsets of the sampled points with the highest $68.27\%,~    95.45\%~\rm{and}~    99.73\%$ fractions respectively of the posterior value among all samples. The different areas of the two different modes reflect the fact that the two modes have different weights. Like all MCMC results, the density of the sampling points reflect the posterior.}
\label{fig:toy}
\end{figure}

In figure \ref{fig:toy} an example of the sampling results is shown, with blue, green and red colour points representing the highest (i.e. largest value of the posterior) $68.27\%,~    95.45\%~\rm{and}~    99.73\%$ fraction of the samples, after sorting the posterior values in descending order.

\section{Discussion}\label{sec:dis}

A novel method, which we call \emph{mixed MCMC}, has been proposed.
In the situations when the multi-modal characteristics of the posterior distribution are already roughly known, the parameter space can be split into several sub-regions, each of which hosts a single mode, and our mixed MCMC method can be applied.  The proposal density can generate candidates in different sub-regions by adding a shift from the current sub-region to the proposed new sub-region. In this way, a comparison between different sub-regions can be done globally, which improves efficiency.  This algorithm is strictly Markovian, so the detailed balance requirement is fulfilled.
The concept of mixed MCMC is realised by enabling proposed candidate points to be generated from different modes of the posterior.
Admittedly, the mixed MCMC approach must rely on other methods to first identify the multiple modes of the posterior distribution.
However, since that identification will generally require only rough information, we can expect this initial stage to be rapid.  Moreover, other existing methods already provide some solutions to the problem of identifying multiple modes\cite{feroz09}\cite{feroz13}.

From another perspective, if we view separate parameter subspaces as different models, this mixed MCMC algorithm can be viewed as an special form of reversible jump MCMC\cite{green95}\cite{green01}, which can sample from different models even when they have different dimensionality, and thus provides the Bayesian odds ratio of two models.

This method is a novel realisation of MCMC which can achieve high efficiency in analysing multi-modal posterior distributions by virtue of its unique form of proposal density.   It relies not only on local information, but also on the global structure through swapping between different sub-chains.  In particular the candidate point is accepted with an \emph{acceptance probability} $a=\min(1,r)$, where \emph{r} is the \emph{Metropolis ratio}, which takes into account the global information about the multiple modes of the posterior.

So far, we have not discussed in detail how to obtain rough information about the posterior modes.
We note that methods such as MultiNest\cite{feroz09}\cite{feroz13} aim to solve similar problems, so their approach could be directly applied here. Some other methods like parallel tempering MCMC\cite{neal96} or {\em k-means\/} \cite{macqueen67} can also be modified and applied here.

We leave the detailed comparison with other methods, like parallel tempering MCMC to future work. However, by not throwing away points in parallel chains, and the design of the proposal density to have a relatively short autocorrelation length we expect the mixed MCMC algorithm to be quite efficient \cite{peskun73}.

In this work, we also applied our method to a simple toy model, with two distinct well-separated modes, to demonstrate its efficacy.
With $10^5$ samples, our mixed MCMC was able to both find the picking up probability, which represents the bulk distribution of the posterior (i.e. the probability of belonging to each mode) and also the Bayesian credible regions for the posterior as a whole -- each of which show excellent agreement with the exact, theoretically computed values for our toy model.

This toy model investigation shows that the idea of mixed MCMC is theoretically sound and practically useful. 
Further investigation can be done in future work by investigating ways of optimising the local proposal densities individually so that the efficiency of the method can be further improved.

\appendix
\section{Analytical Evaluation of $2\Delta\log(\cal L)$}
In this section we present the calculation of $2\Delta\log(\cal L)$, which is defined as $$2\Delta\log({\cal L}) = 2\log( {\cal L} _{max}) - 2\log({\cal L}),$$ for our toy model posterior.

In the toy model, there are two well separated modes, and we can simply assume they are fully independent. The two modes have an evidence ratio of $w_1:w_2$, where $w_1+w_2=1$ is the normalisation requirement.

We define the two independent parts of the posterior as
$$P_1(x,y) = \frac{1}{2\pi\sigma^x_1 \sigma^y_1}\exp[-\frac{(x-\mu^x_1)^2}{2(\sigma^x_1)^2}-\frac{(y-\mu^y_1)^2}{2(\sigma^y_1)^2}]$$
$$P_2(x,y) = \frac{1}{2\pi\sigma^x_2\sigma^y_2}\exp[-\frac{(x-\mu^x_2)^2}{2(\sigma^x_2)^2}-\frac{(y-\mu^y_2)^2}{2(\sigma^y_2)^2}]$$
and the posterior can be written as
\begin{equation}\label{eq:pos}
f(x,y) = w_1P_1(x,y)+w_2P_2(x,y)
\end{equation}
The peak values of the posterior for its two modes are $C_1\triangleq\frac{w_1}{2\pi \sigma_1^x\sigma_1^y}$ and $C_2\triangleq\frac{w_2}{2\pi \sigma_2^x\sigma_2^y}$ respectively.

For simplicity, we replace $\frac{(x-\mu^x_1)^2}{(\sigma^x_1)^2}+\frac{(y-\mu^y_1)^2}{(\sigma^y_1)^2}=r_1^2$ and $\frac{(x-\mu^x_2)^2}{(\sigma^x_2)^2}+\frac{(y-\mu^y_2)^2}{(\sigma^y_2)^2}=r_2^2$ and rewrite the posterior as 
\begin{equation}\label{eq:pos2}
f(x,y) = C_1\exp(-r_1^2/2)+C_2\exp(-r_2^2/2)
\end{equation}

Without losing generality, we assume $C_1>C_2$, and so the highest posterior value $f_{max}=C_1$, and highest posterior value for the secondary peak is $C_2$.
We define $r_0$ as $C_1\exp(-r_0^2/2) = C_2$, equivalently, $\exp(-r_0^2/2) = \frac{C_2}{C_1}$.

Our aim is to find the expression for $\Delta\chi^2(C)$, so that given $C$, we have $$\int\limits_{f>\exp(-\frac{\Delta\chi^2}{2})} f(x,y)~\mathrm{d}x~\mathrm{d}y = C$$

When $f > \exp(-r_0^2/2) $,

\begin{eqnarray}\label{eq:small}
C &=&\int\limits_{f>\exp(-\Delta\chi^2/2)}C_1\exp(-\frac{r^2}{2})\mathrm{d}x~\mathrm{d}y\nonumber\\
&=&2\pi C_1 \sigma_1^x \sigma_1^y\int_0^{\Delta\chi^2}\exp(-\frac{r^2}{2})\mathrm{d}r^2/2\nonumber\\
&=&w_1[1-\exp(-\Delta\chi^2/2)].
\end{eqnarray}
 
This expression is valid so long as $C<C_0 \triangleq w_1(1-\frac{C_1}{C_2})=w_1 - w_2\frac{\sigma_1^x\sigma_1^y}{\sigma_2^x\sigma_2^y}$, $=w_1-w_2$; if, however, $C$ is bigger, than we have to include the contribution from the secondary mode.

\begin{equation}\label{eq:big}
\begin{split}
&C =C_0+\int_{r_0}^{r_1}w_1\exp(-\frac{r^2}{2})r\mathrm{d}r+\int_0^{r_2}w_2\exp(-\frac{r^2}{2})r\mathrm{d}r\\
&C - C_0=w_1[\frac{C_2}{C_1}-\exp(-r_1^2/2)]+w_2[1-\exp(-r_2^2/2)]\\
&C - w_1 + w_2 \frac{\sigma_1^x\sigma_1^y}{\sigma_2^x\sigma_2^y}=w_2\frac{\sigma_1^x\sigma_1^y}{\sigma_2^x\sigma_2^y}-w_1\exp(-r_1^2/2)\\
&~~~~~~+w_2-w_1\frac{\sigma_2^x\sigma_2^y}{\sigma_1^x\sigma_1^y}\exp(-r_1^2/2)\\
&C=1 - w_1(1+\frac{\sigma_2^x\sigma_2^y}{\sigma_1^x\sigma_1^y})\exp(-r_1^2/2)
\end{split}
\end{equation}
In the third line we used the relation that $w_1P(r_1)=w_2P(r_2))$.

Furthermore, we have 
\begin{equation}\label{eq:further}
\begin{split}
\exp(-\frac{r_1^2}{2})&=\frac{1-C}{w_1(1+\frac{\sigma_2^x\sigma_2^y}{\sigma_1^x\sigma_1^y})}\\
-r_1^2/2 &= \log(1-C) - \log(w_1)-\log(1+\frac{\sigma_2^x\sigma_2^y}{\sigma_1^x\sigma_1^y})\\
\Delta\chi^2=r_1^2&=-2[\log(1-C) - \log(w_1)\\
&~~~~~~~-\log(1+\frac{\sigma_2^x\sigma_2^y}{\sigma_1^x\sigma_1^y})]
\end{split}
\end{equation}

We determined $w_1 = \frac{3}{4}$ and $w_2 = \frac{1}{4}$, while keeping $\sigma_1^x = \sigma_2^x$ and $\sigma_1^y = \sigma_2^y$, and choosing the C value as $68.27\%,~    95.45\%~\rm{and}~    99.73\%$.  This yields the corresponding $\Delta\chi^2$ values as 3.11, 6.99 and 12.64.

\bibliography{main}

\end{document}